\documentstyle[preprint,12pt,aps,epsfig]{revtex}

\tighten
\begin{document}
\draft
\preprint{Imperial/TP/95-96/62}

\title{A New Approach to QCD Sum Rules and Inclusive $\tau$ Decay}
\author{H. F. Jones$^a$\footnote{email: h.f.jones@ic.ac.uk}, 
        A. Ritz$^a$\footnote{email: a.ritz@ic.ac.uk}, and 
        I.L. Solovtsov$^{ab}$\footnote{email: solovtso@thsun1.jinr.dubna.su}\\ 
        $\;$\\}
\address{$^a$ Physics Department, Imperial College, \\
         Prince Consort Rd., London, SW7 2BZ, \\
         United Kingdom\\ $\;$\\}
\address{$^b$ Bogoliubov Laboratory of Theoretical Physics, \\
         Joint Institute for Nuclear Research, \\
         Dubna, Moscow Region, 141980,\\ Russia \\ $\;$}
\date{\today}

\maketitle

\begin{abstract}

We show that
renormalization group improvement, making use of analyticity
and the structure of the operator product expansion, combined
with a non-perturbative
expansion method allows one to describe quarkonium states via QCD sum
rules without the need for explicit power corrections. This technique also
allows an accurate determination of parameters related to 
the inclusive decay of the $\tau$ lepton.
\end{abstract}

 

While the QCD sum rules technique \cite{svz79} provides a 
powerful method for extracting hadronic characteristics from QCD, there
is the drawback that a description of the large distance behaviour
via power corrections associated with quark and gluon condensates introduces
extra phenomenological parameters.
In this Letter we shall argue that the non-perturbative expansion technique
of Ref.~\cite{solov94}, based on a new small expansion parameter,
combined with a particular
renormalization group improvement which respects both
analyticity and the structure of the operator product expansion,
enables us to describe some fundamental hadronic properties
without the explicit introduction of condensates.
As a consequence of the retention of the analytic properties of the 
running coupling, which is essentially 
equivalent to introducing a renormalon contribution, one may also evaluate the 
$R_{\tau}$ ratio for the inclusive 
semi-leptonic decay of the $\tau$-lepton both by integration over the physical
region and via use of Cauchy's theorem \cite{jones95}.

The non-perturbative technique we use
involves the so-called ``floating'' or ``variational'' series. In quantum
mechanics this series has been applied successfully to the
anharmonic oscillator \cite{seznec79,halliday}. The variational
series also appears in the linear $\delta$-expansion \cite{duncan93} and in
variational perturbation theory \cite{sissakian94,klein95a}. In the case of
zero- and one-dimensional field theories there exist rigorous proofs of the
convergence of such series \cite{arvan95c,guida95}. 
When one considers QCD, the same ideas lead to an 
expansion for the generating functional in terms of a new 
parameter $a$, related to
the coupling by \cite{solov94}
\begin{equation}
\label{lam}
 \lambda \equiv \frac{g^2}{{(4\pi)}^2}\,=\,\frac{1}{C}\,
            \frac{a^2}{{(1-a)}^3}\, ,
\end{equation}
where $C$ is a positive constant which can be found from meson spectroscopy. 
It is clear that for all values of the 
coupling $\lambda\geq 0$ the expansion parameter $a$ lies in the range
$0\leq a < 1$.

In order to illustrate our technique, let us 
consider the Adler $D$--function $D(Q^2) = -Q^2 d\Pi/dQ^2$
corresponding to the vector hadronic correlator in the massless
case. The two--loop
perturbative approximation is given by $D(t,\lambda)=1+4\lambda(\mu^2)$,
where $t=Q^2/\mu^2$.
Standard renormalization group (RG) improvement leads to the substitution
$\lambda(\mu^2)\longrightarrow\overline{\lambda}(t,\lambda)$, 
which implies a summation
of the leading logarithmic contributions. However, due to the Landau pole 
of the running coupling at $Q^2=\Lambda_{QCD}^2$ this substitution
breaks the analytic properties of the $D$--function in the complex 
 $q^2=-Q^2$ plane, namely that
the $D$--function should only have a cut on the positive real $q^2$ axis.

We may correct this feature by noting that the above
solution of the renormalization group equation is not unique. The general 
solution
is a function of the running coupling with the asymptotic behaviour 
$1+4\lambda$, for small
$\lambda$. To maintain the analytic properties of the 
$D$--function we can write it as the dispersion integral of 
$R(s)=(1/\pi){\rm Im}\Pi(s+i\epsilon)$, namely
\begin{equation}
\label{ddispn}
D(t,\lambda) = Q^2\int_0^\infty \;\frac{ds}{(s+Q^2)^2}
R(s,\lambda),
\end{equation}
and use RG 
improvement on the {\em integrand} rather than $D$ itself.
This method leads to $D(t,\lambda)=1+4\lambda_{{\rm eff}}(t,\lambda)$, 
where, with $\tau = s/Q^2$, 
\begin{equation}
\lambda_{{\rm eff}}=\int_0^\infty \; \frac {d\tau}{(1+\tau)^2}
\frac {\bar\lambda(t,\lambda)}{1 + \bar\lambda(t,\lambda)\beta_0 \ln\tau}.
\end{equation}
This has the Borel representation 
\begin{equation}
\label{Eq.9}
\lambda_{{\rm eff}} (t, \lambda )\,=\,\int_0^\infty\,db\,{\rm e}^
{- b/\overline{\lambda} (t,\lambda)}\,B(b)\;,
\end{equation}
with
\begin{equation}
\label{Eq.10}
B(b)\,=\,\Gamma (1\,+\,b\beta_0) \,\Gamma (1\,-\,b\beta_0)\; .
\end{equation}
Here $\beta_0=11-2/3N_f$ is the first coefficient of the $\beta$-function,
and $N_f$ is the number of active flavours. Thus, in the Borel plane there are
singularities at $b\beta_0=-1,-2,...$ and $b\beta_0=1,2,...$ corresponding to
ultraviolet and infrared (IR) renormalons respectively.

The first IR singularity at $b\beta_0=1$ is probably absent since there
is no corresponding operator in the operator product expansion. 
Although this issue is not currently settled, it seems reasonable
to assume that the first IR renormalon occurs at 
$b=2/\beta_0$, and we would like to use this property of the operator
product expansion as an additional constraint
on the choice of solution to the renormalization group equation.  
This can be achieved by integrating by parts in Eq.~(\ref{ddispn}),
using the fact that to two--loop order $R(s)$ is a constant, and  
applying the same RG improvement to the new integrand,  to obtain the 
following expression for $\lambda_{{\rm eff}}$:
\begin{equation}
\label{Eq.8}
 \lambda_{{\rm eff}} (t,\lambda) = \int_0^\infty\,d\,\tau \, \omega(\tau)
     \frac{\overline{\lambda}(kt,\lambda)}{1+\overline{\lambda}(kt,\lambda)
            \beta_0\, \ln\tau}\,,      
\end{equation}
in which the factor $k$ reflects the renormalization scheme ambiguity and
the function
\begin{equation}
\label{Eq.13}
\omega (\tau )\,\equiv\,\frac{2\,\tau}{{(\,1\,+\,\tau\, )}^3}
\end{equation}
describes the distribution of virtuality usually associated with
renormalon chains. The function (\ref{Eq.13}) coincides with the function
used in \cite{bigi94} and is numerically very close to that
found in \cite{neubert95}. 
The Borel transform of (\ref{Eq.8}) has the form
\begin{equation}
\label{Eq.11}
B(b)\,=\,\Gamma (1\,+\,b\beta_0) \,\Gamma (2\,-\,b\beta_0)\; .
\end{equation}
Thus in this representation for $\lambda_{{\rm eff}}$ 
the positions of all ultraviolet singularities remain unchanged, but 
the first IR renormalon singularity at $b=1/\beta_0$ is absent.

In summary, a representation for the effective coupling, and consequently
the $D$--function, which manifests renormalon--type characteristics 
can be obtained as a particular
RG improvement of the lowest order radiative corrections
which takes into account both the analytic properties and the structure
of the operator product expansion. In order to render Eq.~(\ref{Eq.8})
integrable we must combine this method with the 
non-perturbative $a$-expansion of Ref.~\cite{solov94} 
in which from the beginning the running coupling 
has no ghost pole.
Effectively, the representation for the $D$--function obtained in such
a way coincides with a technique explicitly introducing 
power corrections \cite{pwer}, and we can,
in principle, describe hadronic parameters using, say, the method
of QCD sum rules.

However, retention of the correct analytic properties under RG improvement
also allows for the possibility of considering inclusive $\tau$ decay, and
we shall consider this first. Using RG improvement following the procedure 
described above in the context of the $a$--expansion leads to 
\begin{equation}
\lambda_{{\rm eff}}(q^2) = -q^2\int_0^{\infty} d\sigma 
      \frac{2\sigma}{(\sigma-q^2)^3}  \tilde{\lambda}(\sigma),\label{ltilde}
\end{equation}
where to leading order
\begin{equation}
 \tilde{\lambda} = \frac{a^2}{C}(1+3a) \label{lameff}
\end{equation}
The renormalization group defines the running of $a(\sigma)$ 
as the solution of the transcendental equation \cite{solov94}
\begin{equation}
\label{rga}
  \sigma = \sigma_0 \exp\left[\frac{C}{2\,\beta_0}\, \left(\,
     f(a)\,-\,f(a_0)\, \right)\right],
\end{equation}
where
\begin{eqnarray}
\label{f}
  f(a) & = & \frac{2}{a^2} -   \frac{6}{a} - 48\ln a-
    \frac{18}{11}\,\frac{1}{1-a} \nonumber\\
  & & +  \frac{624}{121}\,\ln{(1-a) } +
    \frac{5184}{121}\,\ln{( 1+\frac{9}{2}\,a )} \, .
\end{eqnarray}
The parameter $a_0$ and the virtuality $\sigma_0$ in Eq.~(\ref{rga}) are 
defined by some renormalization point for the effective coupling.

We now apply the representation (\ref{ltilde}) to inclusive
$\tau$--decay. The starting point is the
expression \cite{jones95}
\begin{equation}
\label{Eq.19}
R_{\tau}\,=\,2\,\int_0^{M_{\tau}^2}\,\frac{ds}{{M_{\tau}^2}}\,
{\left(1\,-\frac{s}{{M_{\tau}^2}}\right)}^2\,
\left( 1\,+\,\frac{2s}{{M_{\tau}^2}}\right)\,\tilde R(s),
\end{equation}
where $\tilde R(s) = R_\tau^0 R(s)$ and
\begin{equation}
\label{Eq.20}
 R_{\tau}^0 = 3\left( |V_{ud}|^2+|V_{us}|^2\right)S_{EW}
\end{equation}
in which the electroweak factor $S_{EW}=1.0194$ and the CKM matrix
elements are $|V_{ud}|=0.9753$, $|V_{us}|=0.221$, taken from 
Ref.~\cite{braaten92}.
Then, in order to isolate the QCD correction to $R_\tau$, we write 
$R_\tau = R_\tau^0\left(1\,+\,\Delta R_{\tau}\right)$.

The effective coupling (\ref{ltilde}) is an analytic function of $s$ in the
complex $s$--plane with a cut along the positive real axis.
 $\Delta R_{\tau}$ may  be written as the contour integral \cite{jones95}
\begin{equation}
\label{Eq.21}
\Delta R_{\tau}\,=\,\frac{d_1}{2\pi {\rm i}}\,
\oint_{|z|=1}\,\frac{dz}{z}\,{(1-z)}^3\,(1+z)\,
\lambda_{{\rm eff}}\,(M_{\tau}^2\,z)\, ,
\end{equation}
where $d_1=4$ is the 2-loop coefficient of the D--function.

Substituting Eq.~(\ref{ltilde}) into Eq.~(\ref{Eq.21}) and
using Cauchy's theorem we obtain
\begin{equation}
\label{Eq.22}
\Delta R_{\tau}\,=\,12d_1\int_0^{M_\tau^2}\frac{ds}{M_\tau^2}
\left(\frac {s}{M_\tau^2}\right)^2\left(1-\frac {s}{M_\tau^2}\right)
\tilde{\lambda}(ks),
\end{equation}
in which the factor $k$ again parametrizes the renormalization scheme. 
In what follows we shall always use the $\overline{MS}$ scheme, 
in which  \cite{neubert96} $k=\exp(-5/3)$. Note that
in comparison with (\ref{Eq.19}), use of the representation (\ref{ltilde})
for the coupling modifies the kinematic factor so that the maximum
now occurs near $s=(2/3)M_{\tau}^2$. 

Taking as input the experimental value of 
$R_{\tau}^{\rm exp}\,=\,3.56\pm 0.03$~\cite{pich95}, three active quark
flavours and the variational parameter $C=4.1$ as in \cite{jones95},
we find
\begin{equation}
\label{Eq.23}
\alpha_s(M_{\tau}^2)\,=\,4\pi\,\tilde\lambda(M_{\tau}^2)\,=\,
0.339\,\pm\,0.015\, .
\end{equation}
which differs significantly from that
obtained ($\alpha_s(M_{\tau}^2)=0.40$ in leading order \cite{jones95}) 
without the renormalon-inspired  representation (\ref{ltilde}) for the 
coupling. The new method, applying the matching procedure in the physical
$s$--channel and using standard heavy quark masses \cite{pdg94}, leads
to $R_Z=20.90\pm 0.03$, which
agrees well with experimental data \cite{pdg94}.

As mentioned above, the requirement of analyticity naturally generates
power corrections. Consequently, it is natural to investigate whether this 
technique allows us to describe meson parameters using the QCD sum rules
approach. Consider, for example, the hadronic correlator $\Pi(s)$ corresponding
to the vector current \cite{va}. 
The imaginary part of $\Pi(s)$ to order $a^3$ in the non-perturbative expansion
of Ref.~\cite{solov94}
is given by 
\begin{equation}
 {\rm Im} \Pi(s) =  \frac{1}{4\pi} \left[\Pi^{(0)}(s)
         +4\tilde\lambda \Pi^{(1)}(s)\right],
\end{equation}
where $\tilde\lambda$ is given by Eq.~(\ref{lameff}) 
and for $\Pi^{(0)}$ and $\Pi^{(1)}$ we use the perturbative two-loop 
expressions given in \cite{rry85}. It should be stressed that while
the $a$-expansion technique makes use of the 
perturbative coefficients, the structure of the new expansion is 
fundamentally different.
That it is indeed valid to use perturbative formulae in
a non-perturbative calculation can be seen by noting that for the
$c\overline{c}$ bound states to be 
considered, the dominant contribution to the moment integrals comes from an
energy scale well above $\Lambda_{QCD}$.

The initial expression for the first power moment is given by
\begin{eqnarray}
 M_{1}(Q^2)\,=\,\frac{D(Q^2)}{Q^2} &\equiv& -\frac{d\Pi(Q^2)}{dQ^2}
                 \nonumber\\
        & = & \frac{1}{\pi} \int_{0}^{\infty} d\sigma
           \frac{{\rm Im} \Pi(u)}{(Q^2+\sigma+4m^2)^2},
\end{eqnarray}
where $u=\surd(\sigma/(\sigma+4m^2))$.
In analogy with the massless case we shall associate the parameter 
$\sigma=s-4m^2$ with virtuality. The zeroth--order contribution is given
by
\begin{equation}
 M_1^{(0)}(Q^2) = \frac{1}{4\pi^2}\int_0^{\infty} d\sigma
       \frac{\Pi^{(0)}(s)}{(Q^2+\sigma+4m^2)^2},
\end{equation}
while the order $O(\tilde\lambda)$ correction term 
associated with
$\Pi^{(1)}$ can be rewritten, again in analogy with the 
massless case, as 
\begin{equation}
 M_1^{(1)}(Q^2) = \frac{\tilde\lambda}{\pi^2} \int_{0}^{\infty} d\sigma
           \frac{2(\sigma+4m^2)}{(Q^2+\sigma+4m^2)^3}\Psi(u),
\end{equation}
where
\begin{equation}
 \Psi(u) = (1-u^2)\int_0^u dv \frac{2v}{(1-v^2)^2}\Pi^{(1)}(v).
\end{equation}
The structure of the integrand in the expression for $M_1$ allows us to 
simulate a summation of all terms by placing the term proportional to
$\tilde\lambda$ in the denominator (cf. the mass correction for the
propagator). As a result we have the expression 
\begin{equation}
 M_1(Q^2) = \frac{1}{4\pi^2} \int_{0}^{\infty} d\sigma
           \frac{\Pi^{(0)}(u)}{(Q^2+W(\sigma))^2},
\end{equation}
where, after RG improvement,
$W(\sigma)$ has the form 
\begin{equation}
 W(\sigma) = (\sigma+4\overline{m}^2(k\sigma))
     \left[1-4\tilde\lambda(k\sigma)\frac{\Psi(u)}{\Pi^{(0)}(u)}
               \right],
\end{equation}
and now $u=\surd(\sigma/(\sigma+4\overline{m}^2(k\sigma)))$.
Note that although this procedure introduces correction terms of order
 $\tilde\lambda^2$, it is valid a posteriori, as the dominant 
contribution
to the moment integral comes from the region 
where $\tilde\lambda$ is small \cite{light}.

Evaluation of this expression uses the running expansion parameter
$\overline{a}(\sigma)$ obtained from Eq.~(\ref{rga}), while the running
mass, $\overline{m}(\sigma) = m_0 
(\overline{\lambda}(\sigma)/\overline{\lambda}(\sigma_0))^{\gamma_0/\beta_0}$, 
has the standard perturbative form, with
$\gamma_0$ the first coefficient of the anomalous dimension. However, in this
case
$\overline{\lambda}(\sigma)$ is expressed via Eq.~(\ref{lam}) in terms of the
running expansion parameter $\overline{a}$. The moments of the
vector correlator for general $n$ are then given by
\begin{equation}
 \label{mom}
 M_n(Q^2) = \frac{1}{4\pi^2} \int_{0}^{\infty} d\sigma
           \frac{\Pi^{(0)}(u)}{(Q^2+W(\sigma))^{n+1}}.
\end{equation}
A crucial feature of this approach is that the function $W(\sigma)$ achieves
a minimum at some $\sigma=\tilde{\sigma}$, and for large $n$ the dominant
contribution comes from this point. Defining the ratio of the moments,
$R_n(Q^2) \equiv M_{n-1}(Q^2)/M_n(Q^2)$, we see that for 
large $n$ this ratio tends asymptotically to its saddle--point approximation,
i.e.
\begin{equation}
 R_n(Q^2)\stackrel{n\rightarrow\infty}{\longrightarrow}Q^2 
           + W(\tilde{\sigma}). \label{rn}
\end{equation}
On the other hand the corresponding phenomenological ratio for mesons, 
 $R_n^{\rm had}(Q^2)$, has the form  $Q^2+M^2$ at large $n$, 
where $M$ is the mass of the first
resonance in the relevant channel. Consequently, we obtain
\begin{equation}
 \label{mass}
 M = \sqrt{W(\tilde{\sigma})}.
\end{equation}
In \cite{svz79} the moments were considered at $Q^2=0$, while
the case of $Q^2\neq 0$ has also been considered in the literature
(see \cite{rry85} for a review). It is an important feature of this approach
that the analysis
is independent of $Q^2$, as the functional form of the $Q^2$-dependence
for large $n$ is the same for both the QCD and phenomenological
moment ratios.

\begin{figure}
 \centerline{
   \psfig{file=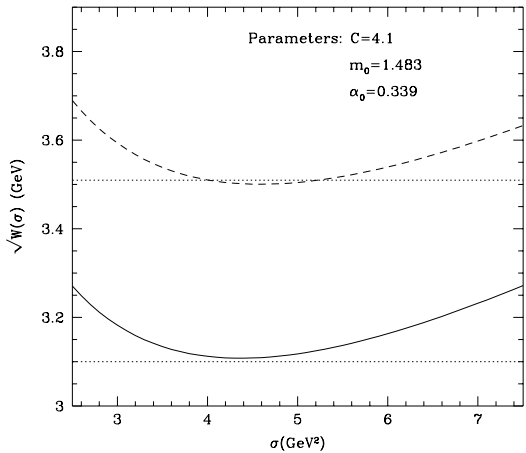,width=7.5cm,angle=0}
   }
 \caption{A plot of $\protect\sqrt{W(\sigma)}$ versus $\sigma$ for the vector
  (solid curve), and axial-vector (dashed curve) currents,
  the minimum being the
  asymptotic limit of the moment ratios for large $n$. For comparison,
  the straight lines are the corresponding
  experimental $c\overline{c}$ bound state masses \protect\cite{pdg94}.} 
  \label{cc}
\end{figure}

Estimates for the $c\overline{c}$ bound state masses
for the vector and axial vector channels, 
corresponding to (\ref{mass}) are shown in Fig.~\ref{cc}
for values of $\sigma$ near $\tilde{\sigma}$. The analysis has been 
presented thus far for the vector channel only. The calculation in the 
axial-vector case is similar, the modifications to the moment expressions
being standard, and will not be discussed here (see e.g. \cite{rry85}).
Keeping $C=4.1$ and taking $\alpha_0$  at the $\tau$
mass scale from Eq.~(\ref{Eq.23}) we have only one free parameter, 
the quark mass $m_0$. With $m_0=1.483$ GeV we obtain a good fit to the
experimental masses \cite{pdg94} of both $J/\psi(1S)$ and $\chi_{c1}(1P)$, 
as shown in Fig.~\ref{cc}.

Note that the parameter $C=4.1$ used here was found \cite{solov94} from
a fit to the $\beta$-function and corresponds to the linear growth of the 
quark-antiquark potential. The present calculation is consistent with
this derivation, since the dominant contribution to the moment integrals 
arises from a region where the system is still reasonably non--relativistic
($u^2 \sim 0.3$), 
 
In conclusion, we have presented in this Letter 
a technique for obtaining quarkonium and $\tau$--decay
parameters which makes use of RG improvement and a non-perturbative
expansion which removes the Landau pole in the running coupling.
It appears that careful control of RG 
improvement, to ensure the correct analytic properties, naturally induces
power corrections \cite{shirkov96} required for the description of 
meson parameters
without the need for their explicit
introduction. In the case of QCD sum rules the fact that 
this technique involves only three parameters, two
being fundamental to QCD, i.e. the coupling constant $\alpha_0$ and the 
quark mass $m_0$ defined at some energy scale, suggests that it is applicable
to the study of other channels in the $c\overline{c}$ family and also
to consideration of other heavy quarkonium states. 
Such an extension is currently under investigation and the
results will be published elsewhere \cite{jones96d}.

We thank D. Ebert, D.I. Kazakov, S.V. Mikhailov, A.N. Sissakian,
O.P. Solovtsova, and A.A. Vladimirov for interest in the work and 
useful comments. The financial support of A.R. by the
Commonwealth Scholarship
Commission in the U.K. and the British Council, and of I. L. S.
by the Royal Society and RFBR (grant 96-02-16126-a) is gratefully acknowledged.
I.L.S. also thanks Prof. T.W.B. Kibble and the Theoretical Physics
Group for their warm hospitality at Imperial College. 

\bibliographystyle{prsty}

\begin{thebibliography}{10}
\bibitem{pwer}
Recently \cite{shirkov96}, it has been shown that requiring
the correct analytic
properties for the running coupling is indeed equivalent to the inclusion
of  non-perturbative
power corrections of the form $\exp(-1/(\overline{\lambda}(Q^2)\beta_0))$.
\bibitem{va}
We shall only consider the vector 
and axial-vector
currents, for which there are reliable two-loop expressions.
\bibitem{light}
For  mesons constructed from light quarks the main contribution to the 
moment integrals would come from the
region of large $\overline{a}$, near its limiting value of unity. In order to
consider these systems the role of next to leading order corrections
would need to be addressed.
\end{thebibliography}

\end{document}